\documentclass[twocolumn,prl,showpacs,superscriptaddress,preprintnumbers,amsmath,amssymb]{revtex4}

\usepackage[dvips]{graphicx}
\usepackage{dcolumn,color}
\usepackage{bm}
\begin{document}

\title{Jahn-Teller distortions and charge, orbital and magnetic orders in NaMn$_7$O$_{12}$}

\author{Sergey V. Streltsov}
\affiliation{Institute of Metal Physics, S.Kovalevskoy St. 18, 620990 
Ekaterinburg Russia}
\affiliation{Ural Federal University, Mira St. 19, 620002 
Ekaterinburg, Russia}
\email{streltsov@imp.uran.ru}

\author{Daniel I. Khomskii}
\affiliation{II. Physikalisches Institut, Universit$\ddot a$t zu K$\ddot o$ln,
Z$\ddot u$lpicher Stra$\ss$e 77, D-50937 K$\ddot o$ln, Germany}

\begin{abstract}
With the use of the band structure calculations we demonstrate that previously
reported [Nat. Materials {\bf 3}, 48 (2004)] experimental crystal and magnetic structures for 
NaMn$_7$O$_{12}$ are inconsistent with each other. The optimization of the crystal lattice 
allows us to predict a new crystal structure for the low temperature phase, which is 
qualitatively different from the one presented before. The AFM-CE type of the magnetic order 
stabilizes the structure with the elongated, not compressed Mn$^{3+}_B$O$_6$ octahedra, 
striking NaMn$_7$O$_{12}$ out of the list of the anomalous Jahn-Teller systems. The
orbital correlations were shown to exist even in the cubic phase, while
the charge order appears only in the low temperature distorted phase.
\end{abstract}

\maketitle

{\it Introduction.--} The quadruple perovskites based on
transition metal (TM) ions with general formula A'A''$_3$B$_4$O$_{12}$ 
and mixed occupation of the A sites of initial ABO$_3$ perovskite structure by
A' (cations with large ionic radii A' = La, Na, Ca, Bi etc.), 
A'' (Jahn-Teller (JT) ions like Mn$^{3+}$, Cu$^{2+}$) ions are known for 
unusual physical properties. CaCu$_3$Ti$_4$O$_{12}$ shows a giant dielectric 
constant,~\cite{Homes2001} LaCu$_3$Fe$_4$O$_{12}$ demonstrates nontrivial charge 
ordering,~\cite{Long2009} CaMn$_7$O$_{12}$ is multiferroic with the largest magnetically 
induced electric polarization measured to date,~\cite{Johnson2012} associated with 
incommensurate structural modulations coupled with the rotation of the singly occupied $e_g$ 
orbital of the Mn$^{3+}$ ion.~\cite{Perks2012} 
The ferroelectric properties of another quadruple perovskite BiMn$_7$O$_{12}$ are also 
related to the orbital degrees of freedom, which is justified by the
study of different doping regimes.~\cite{Mezzadri2011} 

The A site ordered quadruple perovskite NaMn$_3$Mn$_4$O$_{12}$ was shown to 
have intriguing properties. The authors of Ref.~\onlinecite{Prodi2004} 
argued that this system, which on octahedral B sites contains Mn ions with the 
average valence 3.5+, could be similar to half-doped manganites 
like La$_{0.5}$Ca$_{0.5}$MnO$_3$, and is better than the latter because it does 
not show disorder caused in the usual manganites by doping.  Thus one could 
hope to get the ``cleanest'' signatures of charge and orbital order
typical for half-doped manganites. And indeed, they discovered such ordering, 
occurring in NaMn$_7$O$_{12}$ at 180 K, but of completely different type. 
With further decrease of the temperature authors of Ref.~\onlinecite{Prodi2004}
observed the same magnetic structure of the CE type (zigzag 
chains in the $ac$ plane) as in the usual half-doped
manganites.
 
The occupied orbitals of the JT Mn$^{3+}$ ions in 
half-doped manganites are of $3x^2-r^2$ and $3y^2-r^2$ type, forming 
stripes in the basal plane, with locally elongated MnO$_6$ octahedra, 
long axes alternating in the $x-$ and $y-$directions. In contrast, it was argued
in Ref.~\onlinecite{Prodi2004} 
that in NaMn$_7$O$_{12}$ there exist local compression of these octahedra, 
with occupied orbitals being $x^2-y^2$. This however is very surprising. 
Local compression around JT ions with $e_g$ degeneracy in insulators is extremely 
rare: among hundreds of  known such systems there is practically none, or at best very few, 
examples with compressed octahedra. This is mostly due to the anharmonicity 
of the elastic interaction, and to higher-order Jahn-Teller 
coupling.~\cite{Kanamori1960,KhomskiiBrink2000,Bersuker2006}

Thus, the compounds with compressed Jahn-Teller octahedra for the case of $e_g$ degeneracy
is a very rare phenomenon, and there must be special reasons for such distortions - for example 
the layered structure, as in La$_{0.5}$Sr$_{1.5}$MnO$_4$. But even in this cases 
the occupied orbitals are typically of 
$3x^2-r^2$ and $3y^2-r^2$ type, not $x^2-y^2$.~\cite{Wu2011}
Thus this conclusion of Ref.~\onlinecite{Prodi2004} is extremely unusual, and it could be 
very important to understand the reasons of such a behavior. This could have a fundamental 
importance for the large class of materials and phenomena, connected with the behavior of 
systems with JT ions and with orbital ordering - phenomena which play more and more 
important role in modern solid state physics.~\cite{Tokura2000}

In the present paper magnetic, electronic and structural properties of 
NaMn$_7$O$_{12}$ were studied using $ab$ $initio$ band structure
calculations. We show that the charge ordering indeed occurs in the 
low temperature phase, but experimentally claimed~\cite{Prodi2004} magnetic and crystal structures are 
inconsistent with each other. Performing the optimization
of the atomic positions and unit cell vectors with fixed volume 
we found a new crystal structure consisting of the elongated MnO$_6$ octahedra.
The orbital correlations resulting in such a structure are shown to exist
even in a nondistorted cubic lattice. Thus, NaMn$_7$O$_{12}$ is not an exception from the 
general rule of only elongated octahedra for $e_g$ degenerate Jahn-Teller ions, 
formulated above.

{\it Crystal and magnetic structure.--} 
It is worthwhile to rewrite the chemical 
formula of this compound as (NaMn$^{3+}_3$)(Mn$^{3+}_2$Mn$^{4+}_2$)O$_{12}$,
which shows the relationship to the perovskite structure ABO$_3$. Thus a part of the Mn$^{3+}$ 
($d^4$) ions occupies A positions (Mn$^{3+}_A$) in initial perovskite structure having square surrounding, 
while the rest of the Mn ions are situated in the octahedral 
B sites (Mn$^{3+}_B$ and Mn$^{4+}_B$). The checkerboard in the $ac$ plane charge order 
of Mn$_B$ develops below $\sim$180~K (see Fig. 1 in Ref.~\onlinecite{Prodi2004}).
Mn chains of the same valence are formed along the $b$ axis.

Above transition to the charge ordered state NaMn$_7$O$_{12}$ is paramagnetic (PM)
and does not show long range magnetic order down to T=125~K, when the spins of the octahedral 
Mn form AFM-CE type structure~\cite{Prodi2004,Gauzzi2005}. With further decrease
of the temperature below 90~K also the Mn$^{3+}_A$ ions order in the 
antibody-centered AFM state.

{\it Details of the calculations.--} Crystallographic data used in the calculations were taken 
from the Ref.~\onlinecite{Prodi2004}. We used the Linear muffin-tin orbitals (LMTO) 
method~\cite{Andersen1984} for the calculation in the experimental structure. The relaxation 
of this structure was performed using pseudopotential (PP) Vienna ab 
initio simulation package (VASP)~\cite{Kresse1996} in the frameworks
of the projector augmented wave  method~\cite{Blochl1994}.

In spite of the fact that both methods (LMTO and PP) are not
full-potential, they were successfully used previously for the
study of the Jahn-Teller effects~\cite{Binggeli2004,Trimarchi2005,Medvedeva2002,Streltsov12Cu}. 
The potential in the LMTO is spherical, but kinetic
part retains the symmetry of the lattice giving rise to
appropriate orbital pattern as it will be shown below.

The von Barth-Hedin~\cite{Barth1972} and Perdew-Burke-Ernzerhof (PBE)~\cite{Perdew1996} 
versions of the exchange-correlation potentials were utilized in the LMTO and PP calculations
respectively. The strong Coulomb correlations were taken into account via the LSDA+U (for LMTO) and 
GGA+U (for PP) methods.~\cite{Anisimov1997} The values of on-cite Coulomb interaction 
($U$) and Hund's rule coupling ($J_H$) parameters were taken to $U$=4.5 eV 
$J_H$=0.9 eV.~\cite{Streltsov2008} The integration in the course of the
self-consistency was performed over a mesh of 
144 {\bf k}-points in the irreducible part of the Brillouin-zone.
The ionic relaxation was performed using  conjugate-gradient algorithm
with convergence criteria for the total energy 10$^{-4}$ eV.
\begin{figure}[t!]
 \centering
 \includegraphics[clip=false,width=0.4\textwidth]{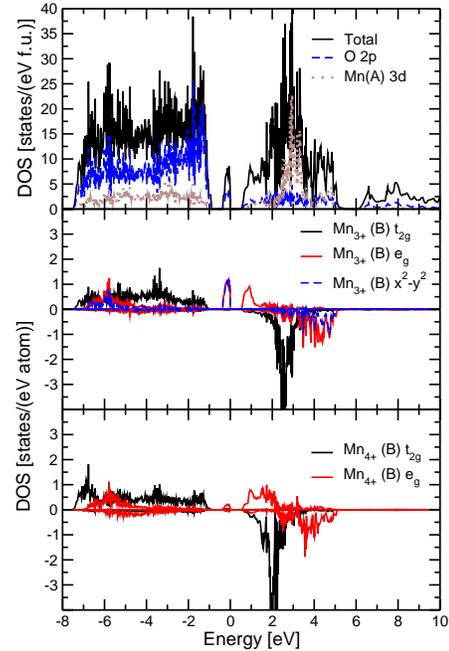}
\caption{\label{DOS-LT}(color online). 
Results of the LSDA+U calculation (LMTO) for the low temperature experimental
crystal structure (compressed Mn$^{3+}$O$_6$ octahedra): total and partial 
densities of states (DOS). AFM-CE type of the magnetic structure was used. 
Fermi energy is in zero.} 
\end{figure}

We utilized Lichtenstein's exchange interaction parameter (LEIP) calculation 
procedure~\cite{LEIP1} for the Heisenberg Hamiltonian, which is written as
$H = \sum_{ij}J_{ij} \vec{S_i} \vec{S_j}$. Summation here runs twice over 
every pair $i,j$.

{\it Low temperature phase, experimental structure.--} 
We start the investigation of NaMn$_7$O$_{12}$ with the LSDA+U calculations of the 
low temperature (LT) experimental
structure. The total and partial densities of states (DOS) obtained for the 
AFM-CE type of magnetic order  are shown in Fig.~\ref{DOS-LT}. It is easy to see that 
indeed two octahedral Mn ions show different charge states,
which agrees with experimental expectations.~\cite{Prodi2004}

One class of the octahedral Mn$_B$ ions has $t^{3}_{2g}e_g^1$ (i.e. Mn$^{3+}_B$) 
configuration with the 
half-filled $x^2-y^2$ orbital (middle panel in Fig.~\ref{DOS-LT}). The top of the valence band is 
defined exactly by these states. 
This in turn agrees with the crystal field theory: the compression of the MnO$_6$
octahedra along local $z-$axis should result in the crystal field splitting of the 
$e_g$ shell, such that the $3z^2 - r^2$ goes higher in energy and an electron localizes on the 
$x^2 - y^2$ orbital. The detailed analysis of the occupation matrices shows that the occupied orbital
is the same for all octahedral Mn$^{3+}_B$ ions. The local magnetic moment on this Mn
equals 3.4 $\mu_B$. It is reduced with respect to ionic value (4~$\mu_B$) due to
hybridization with O $2p$ states, which is clearly seen in the DOS plot in the range 
from -7 to -5~eV (most pronounced for the $e_g$ states.) 

Another class of the octahedral Mn ions shows valence state 4+ with 
basically empty $e_g$ shell (there is however some occupation of these states due to 
hybridization with oxygen), see lowest panel in Fig.~\ref{DOS-LT}. 
The magnetic moment on this ion equals 2.9 $\mu_B$.
\begin{figure}[t!]
 \centering
 \includegraphics[clip=false,width=0.45\textwidth]{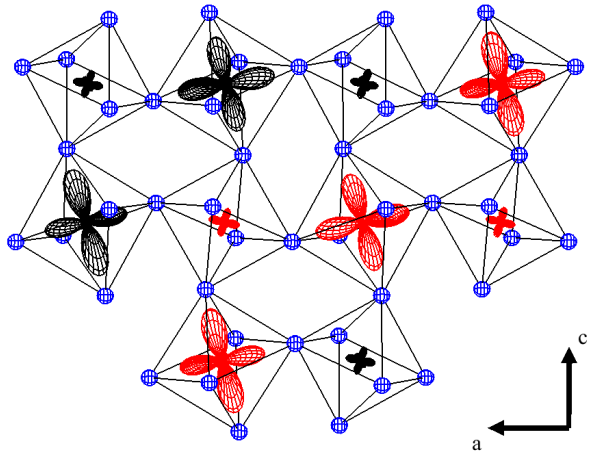}
\caption{\label{OO-LT}(color online). 
Orbital order in the octahedral Mn$_B$ sublattice  
realized in the LSDA+U  (LMTO) calculations of the low temperature  phase 
with experimental distortions (compressed MnO$_6$ octahedra). 
The occupied $e_g$ orbitals are shown.
Magnetic order was chosen to be of AFM-CE type.  Black and red orbitals correspond to 
different spins. Blue balls are oxygens. The Mn$^{4+}_B$ ions are sitting
in the corners of the zig-zag chains, while Mn$^{3+}_B$ in the middle
of the zig-zag bars.} 
\end{figure}

The orbital ordering obtained in the distorted LT phase is presented in Fig.~\ref{OO-LT}. 
One may see, that indeed the single electron in the $e_g$ shell 
of Mn$^{3+}_B$ ions is stabilized on the $x^2-y^2$ orbital.
Since the orbital is the same on each Mn$^{3+}_B$, one may expect that 
the exchange interaction in the $ac$ plane will be FM (exchange between half-filled 
and empty $e_g$ orbitals of Mn$^{3+}_B$ and Mn$^{4+}_B$ respectively)~\cite{Goodenough} with 
{\it all} neighboring Mn$_B$, not only with those forming zig-zag chain. Thus,
this orbital order should stabilize FM, not AFM-CE type of magnetic
ordering in the $ac$ plane.

In order to check this proposal we calculated the exchange constants
with the use of the LEIP formalism. This method allows not only to find
all exchange parameters in one magnetic calculation, but also to analyze,
how stable is a given magnetic configuration (see e.g. Ref.~\onlinecite{Pchelkina2013}). 
The exchange constants between the octahedral Mn$^{3+}_B$ - Mn$^{4+}_B$ ions  
within one zig-zag chain was found to be $J_1 \sim -7 \div 10$~K, all FM.
For the AFM-CE type of the magnetic order the coupling between chains, $J_2$, is expected to 
be AFM, but LEIP calculation shows that $J_2$ could be AFM or FM,
depending on particular pair: $|J_2| \sim 3 \div 6$~K. Thus this magnetic structure 
is unstable within LEIP.~\cite{Pchelkina2013} The total energy
calculations indeed show that e.g. AFM-A$_{ac}$ (ferromagnetic $ac$ planes) solution is 
lower in energy than AFM-CE (on 22 meV/f.u.).
 
To sum up in contrast to naive expectations the experimental LT crystal structure 
leads to the orbital ordering inconsistent with the experimental magnetic order,
the real band structure calculations also show that AFM-CE type of the magnetic
structure does not corresponds to the lowest total energy.

\begin{figure}[t!]
 \centering
 \includegraphics[clip=false,width=0.45\textwidth]{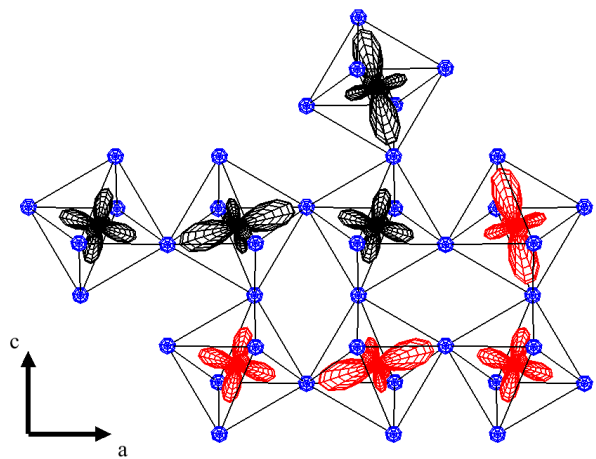}
\caption{\label{OO-HT}(color online). 
Orbital order in the octahedral Mn$_B$ sublattice 
realized in the LSDA+U (LMTO) calculations of the high temperature, cubic 
phase (prior to the lattice optimization). Magnetic order was 
chosen to be AFM-CE type. The occupied $e_g$ orbitals are shown.
Color coding is the same as in Fig.~\ref{OO-LT}.} 
\end{figure}

{\it Orbital order in cubic structure.--}
In order to resolve this inconsistency we performed the calculation
for the cubic structure, taken from the HT phase, but with the cell 
volume corresponding to the LT one. The magnetic structure is experimental, AFM-CE in
the $ac$ plane for the Mn$_B$ and antibody-centered AFM
for Mn$_A$. The orbital ordering in the $e_g$ shell obtained is shown in 
Fig.~\ref{OO-HT}. One may see two distinct 
features different from the results for the experimental LT crystal 
structure (i.e. with Fig.~\ref{OO-LT}).

First of all, there is no clear charge order in this structure,
which is seen from the substantial charge density on the corners of the
the zig-zags, the Mn$^{4+}_B$ ions in the
LT experimental structure. This is related with equal volumes of all
oxygen octahedra surrounding Mn ions sitting in the B sites in this
cubic structure.

The second difference is more important. Even in the absence of the
corresponding distortions of the MnO$_6$ octahedra the single half-filled 
orbital in the $e_g$ shell of the Mn$^{3+}_B$ ions is of the $3z^2-r^2$ symmetry
(actually these are alternating $3x^2-r^2$ and $3y^2-r^2$ orbitals).
Thus the exchange coupling alone (without lattice distortions) stabilizes a 
certain orbital order. This orbital pattern is fully consistent with the 
Goodenough-Kanamori-Anderson rules.~\cite{Goodenough} It explains the ferromagnetic (FM) coupling 
in the zig-zag chains and antiferromagnetic (AFM) between them (due to the half-filled 
$t_{2g}$ orbitals). 

However, stabilization of the $3z^2-r^2$ 
orbital should lead to a certain distortion of the Mn$^{3+}_B$O$_6$ octahedra:
elongation in the $ac$ plane. In order to check this hypothesis we performed
the optimization of the atomic positions and cell shape keeping
the volume constant and equal to the volume of the unit cell in the 
LT phase. The calculations were performed using VASP code within 
the GGA+U approximation. 

{\it Lattice optimization.--}
The Mn$^{3+}_B$O$_6$ octahedra in the optimized structure for the AFM-CE 
type indeed turn out to be elongated, 
not compressed (corresponding crystal structure data are given
in the Supplemental Materials~\cite{Streltsov14sup}). There are two long (2.08 \AA) and four short (two 1.93 \AA~and two 1.92 \AA) 
Mn$^{3+}_B$-O bonds. The Mn$^{4+}_B$O$_6$ octahedra stays to be 
slightly distorted (Mn-O bond lengths are 1.92$\times$2, 1.93, 1.91, 1.88, and 
1.87~\AA). The total energy of the optimized structure for the AFM-CE type of magnetic order
is on 81 meV/f.u. lower than experimental one.

The magnetic moments on two types of the Mn ions sitting in the $B$ sites 
(in the middle of the bar and in the corner of the zig-zags) were found to be 3.6 and 
3.0~$\mu_B$, which certifies the presence of the charge ordered state. The
band gap equals $\sim$0.7 eV.

As it was discussed above for AFM-CE 
the stabilization of the $3x^2-r^2$ and $3y^2-r^2$ orbitals ordering is favorable. It 
coexists with the lattice distortion with locally elongated Mn$^{3+}_B$O$_6$ octahedra.
The mechanism of such ordering could be superexchange interaction,~\cite{KK-UFN}
but it could be also elastic interaction of locally distorted Jahn-Teller centers.~\cite{Sboychakov2011}

Since the strength of the superexchange interaction depends on the
on-site Hubbard repulsion we repeated lattice relaxation
for much larger value of $U=8$~eV and found that 
even for this $U$ octahedra surrounding Mn$_B^{3+}$ ions are
elongated in the optimized structure.

However, there may exist other types of the magnetic ordering
in the $ac$ plane for Mn$_B$, which may result in different orbital pattern and 
different lattice distortions (this occurs e.g. for AFM-A$_{ac}$). 
To check this possibility we carried out the 
crystal structure optimization for
the AFM-A$_{ac}$, AFM-C$_{ac}$ (FM chains in the $ac$ plane), and
AFM-G$_{ac}$ (all neighbors in the $ac$ plane are AFM). 
One may see from Tab.~\ref{total_energies} that these three types of the magnetic structure are higher
in energy than the experimental AFM-CE type.

\begin{table}
\centering \caption{\label{total_energies} Total energies and absolute values of 
spin moments on the Mn ions for different magnetic configurations in 
the $ac$ plane. The atomic positions and the unit cell shape
were relaxed for each magnetic structure in the GGA+U calculation.
Experimental magnetic order for the Mn$_A$ ions and AFM order of spins for Mn$_B$ 
in the $b$ direction were used. We checked that for FM coupling along the $b$ axis
AFM-CE type of also provides the lowest total energy.}
\vspace{0.2cm}
\begin{tabular}{lcccc}
\hline
\hline
     & Total energy & \multicolumn{3}{c}{$|$Magnetic moments$|$ ($\mu_B$)}\\
     & (meV/f.u.)   & Mn$^{3+}_A$ &   Mn$^{3+}_B$ &  Mn$^{4+}_B$ \\
\hline
AFM-CE        & 0   & 3.8 & 3.6 & 3.0\\
AFM-A$_{ac}$  & 58  & 3.8 & 3.7 & 3.1\\
AFM-C$_{ac}$  & 72  & 3.8 & 3.6 & 2.9\\
AFM-G$_{ac}$  & 123 & 3.8 & 3.7 & 2.8\\
\hline
\hline
\end{tabular}
\end{table}

The situation in NaMn$_7$O$_{12}$ reminds that in K$_2$CuF$_4$, in which on the basis of net 
tetragonal compression with $c/a<1$ it was initially concluded that the CuF$_6$ octahedra 
are compressed in the $c$ direction, so that K$_2$CuF$_4$ was even cited in the textbooks as the 
only example with Cu$^{2+}$ in compressed octahedra.~\cite{Goodenough,Ballhausen1962} But it was 
later shown theoretically~\cite{Khomskii1973} and confirmed experimentally~\cite{Ito1976} that  
actually CuF$_6$ octahedra are elongated, but with long axes oriented in the $a$ and 
$b$ directions. 

{\it Discussion.--}
Our results demonstrate that the situation in the ordered phase of 
NaMn$_7$O$_{12}$ should be different from that deduced in Ref.~\onlinecite{Prodi2004}: whereas the observed 
charge and magnetic ordering of the CE type are reproduced in our calculation, the orbital 
order obtained theoretically is completely different from the one previously proposed.
Instead of the $x^2-y^2$ orbitals, occupied at all Mn$^{3+}_B$ ions, we obtained that the 
$3x^2-r^2$  and  $3y^2-r^2$ orbitals should be occupied, forming stripe pattern of the same 
type as in the more conventional half-doped manganites like La$_{0.5}$Ca$_{0.5}$MnO$_3$. 

Such an orbital order naturally explains the AFM-CE type of the magnetic structure, observed experimentally, whereas 
the $x^2-y^2$ orbital ordering proposed in Ref.~\onlinecite{Prodi2004} would give ferromagnetic 
$ac$ planes. The orbital order obtained in the present paper is accompanied (or is caused by) corresponding 
changes of the crystal lattice with not compressed, but elongated  MnO$_6$ octahedra, 
with long axes alternating in $x$ and $y$ (i.e. $a$ and $c$) directions. The 
{\it average} distortion in this case is also a contraction of the unit cell in the $b$ 
direction, as it was found experimentally,~\cite{Prodi2004} but with this average contraction 
being not due to respective compression of the MnO$_6$ octahedra along $b$, but 
rather elongation in the $a$ and $c$ directions.
More careful structural studies, as well as e.g. NMR or XAS measurements 
are expected to reveal corresponding extra lattice distortions predicted 
by the present calculations.

In effect it turns  out that the general rule that the local distortions 
around Jahn-Teller centers with double degeneracy always correspond to local elongation, 
is also valid in this system, so that it seems to be valid  
without any known exceptions. This general message should be kept in mind in studying other 
systems with double $e_g$ orbital degeneracy.


{\it Acknowledgments.--}
S.S. is grateful to M. Korotin, who showed us long ago his results
on the study of the charge and spin states of the Mn ions in this materials
and to A. Gubkin and E. Sherstobitova for useful discussions about the symmetry
of the crystal lattice of NaMn$_7$O$_{12}$. 
This work is supported by the Russian Foundation for Basic Research
via  RFFI-13-02-00374, RFFI-13-02-00050,  the Ministry of 
education and science of Russia  (MK-3443.2013.2). 
The part of the calculations were performed on the ``Uran'' cluster of 
the IMM UB RAS.

\bibliography{../library}
\end{document}